\documentclass[letterpaper]{iopart}
\usepackage{cite}
\usepackage{amsmath,amssymb}
\usepackage{mathrsfs}

\let\ring\mathring

\newcommand\comm[1]{}

\newcommand\topic[1]{\section{#1}}\def\bibpage{\section*{References}}					

\newcommand\defn[1]{\textsl{#1}}

\newcommand\ed{\mathrm{d}}
\newcommand\ee{\mathrm{e}}
\newcommand\ii{\mathrm{i}}
\newcommand\Lie{\mathscr{L}}
\renewcommand\Re{\mathbb{R}}

\newcommand\Hil[1][\relax]{\ifx#1\relax\mathcal{H}\else\HilScr#1\relax\fi}
\def\HilScr#1#2\relax{\Hil#1{\mathrm{#2}}}

\newcommand\grad{\nabla}

\newcommand\idual[1][3mu]{{}^{\star}\mspace{-#1}}

\newcommand\sdual{\mathord{\ast}}

\DeclareMathOperator\real{Re}
\DeclareMathOperator\imag{Im}
\DeclareMathOperator\Int{int}

\newcommand\bip{\beta}
\newcommand\biA{{}^\beta\mspace{-6mu}A}
\newcommand\biom{{}^\beta\mspace{-2mu}\omega}
\newcommand\biOm{{}^\beta\mspace{-1mu}\Omega}

\newcommand\ad[1][\relax]{#1{a}_{\scriptscriptstyle\mspace{-2mu}\Delta}}

\newcommand\Sd[1][\relax]{#1{S}_{\scriptscriptstyle\mspace{-3mu}\Delta}}

\newcommand\Lap[1][\relax]{\ifx#1\relax  \triangle  \else  \triangle_{\mathrm{#1}}  \fi}

\newcommand\expect[2][\relax]{\ifx#1\relax\langle\else\mathopen#1\langle\fi #2 \ifx#1\relax\rangle\else\mathclose#1\rangle\fi}

\newcommand\sR{\mathcal{R}}
\newcommand\sptm{\mathcal{M}}

\makeatletter
\def\smallunderarrow@#1#2#3#4{%
	\vtop{\ialign{##\crcr
		\setbox0=\hbox{$\scriptscriptstyle #1$}%
		\hskip0.5\wd0$\m@th\hfil#3#4$\crcr
		\noalign{\nointerlineskip\kern1.3\ex@}%
		$\scriptscriptstyle #1$\ifx#3\displaystyle\scriptsize\else\tiny\fi#2#3\crcr}}}
\newcommand\pback[1][\relax]{\mathpalette{\smallunderarrow@{#1}\leftarrowfill@}}
\newcommand\ppback[1][\relax]{\mathpalette{\smallunderarrow@{#1}\Leftarrowfill@}}
\makeatother

\begin{document}

\title{Generic isolated horizons in loop quantum gravity}

\author{Christopher Beetle$^1$ and Jonathan Engle$^{1, 2, 3}$}

\address{$^1$ Department of Physics, Florida Atlantic University, 777 Glades Road, Boca Raton, Florida 33431}

\address{$^2$ Centre de Physique Th\'eorique\relax
	\footnote{Unit\'e Mixte de Recherche (UMR 6207) du CNRS et des Universit\'es Aix-Marseille I,
		Aix-Marseille II, et du Sud Toulon-Var;
		laboratoire afili\'e \`a la FRUMAM (FR 2291)},
	Campus de Luminy, Case 907, 13288 Marseille, France}
	
\address{$^3$ Institut f.\ Theoretische Physik III, Universit\"{a}t Erlangen-N\"{u}rnberg, Staudstra{\ss}e 7, 91058 Erlangen, Germany}


\ead{cbeetle@physics.fau.edu, engle@theorie3.physik.uni-erlangen.de}

\bigskip
\begin{indented}
\item 16 July 2010 
\end{indented}

\begin{abstract}
Isolated horizons model equilibrium states of classical black holes.  A detailed quantization,
starting from a classical phase space restricted to spherically symmetric horizons, exists
in the literature and has since been extended to axisymmetry. This paper extends the quantum theory to horizons of arbitrary shape.  Surprisingly, the Hilbert space obtained by quantizing the full phase space of \textit{all} generic horizons with a fixed area is identical to that originally found in spherical symmetry.  The entropy of a large horizon remains one quarter its area, with the Barbero--Immirzi parameter retaining its value from symmetric analyses.  These results suggest a reinterpretation of the intrinsic quantum geometry of the horizon surface.
\end{abstract}



\maketitle

The physics of black holes has opened a fruitful window on quantum gravity.  Singularity theorems \cite{HE:lsst} indicate that black holes form generically in general relativity and contain singularities at which the classical theory breaks down.  Semi-classical arguments \cite{W:bht} bespeak a thermodynamics of black holes whose underlying statistical mechanics is inherently quantum mechanical.  This connection to statistical physics has become a critical testbed for candidate theories of quantum gravity.  In particular, much effort has been devoted to deriving the Bekenstein--Hawking entropy $\Sd = \ad / 4 G \hbar$, where $\ad$ denotes the horizon area, from first principles both in string theory \cite{P:bfst} and in loop quantum gravity \cite{AL:bisr}, to cite but two examples.

This paper concerns the entropy formula in loop quantum gravity.  It builds on work \cite{ABCK:bhe, ACK:ihps, ABK:bhe} initially restricted to black holes with spherically symmetric horizon geometry (see also \cite{S:bhe, R:bhe, K:bhe} before it).  That work is founded on the classical notion of an isolated horizon \cite{AK:idh}, a surface in spacetime whose \textit{quasi-local} definition captures the character of the event horizon in a globally stationary black hole spacetime, but allows for time-dependent fields away from the horizon.  The horizon itself is spherical in \cite{ABCK:bhe, ACK:ihps, ABK:bhe}, but the rest of spacetime is not.  The theory of classical spacetimes with inner boundary at such a horizon admits a consistent canonical formulation \cite{Aea:ihLett, AFK:ham, ABL:mech, ABL:geom}. The symplectic structure in this formulation acquires a horizon surface term that has a form familiar from Chern--Simons theory.  The entropy formula emerges in the subsequent quantization \cite{ABCK:bhe, ACK:ihps, ABK:bhe}.  This pioneering result in spherical symmetry has since been extended to axisymmetry \cite{AEV:qaxi}.

The present paper extends the results of \cite{ABCK:bhe, ACK:ihps, ABK:bhe, AEV:qaxi} still further to \textit{generic} horizon geometries.  The extension yields two surprises.  First, it is possible to quantize the \textit{full} phase space of \textit{all} isolated horizons of a given area without having to fix a sector corresponding to a particular horizon \textit{shape}.  Second, the resulting Hilbert space is \textit{identical} to that found previously.  No new quantum states reflect the broader class of classical horizon geometries allowed prior to quantization.

These facts resolve a tension in \cite{ABCK:bhe, ACK:ihps, ABK:bhe, AEV:qaxi}.  Namely, those analyses consider only symmetric horizons classically, but no analogous symmetry is clearly evident after quantization.  Indeed, spin-networks at the horizon in these quantum frameworks are much like those at other surfaces. The set of fluxes at the horizon, in which quantum geometric information \textit{generically} inheres, is very much the same as at any other surface.  We will discuss these issues in more detail below.  As we shall argue, it is most natural to conclude that \cite{ABCK:bhe, ACK:ihps, ABK:bhe, AEV:qaxi} \textit{actually} quantize the phase space of horizons unrestricted by symmetry presented in this paper.  In this sense, the present work renders the definitive quantization \cite{ABK:bhe} of isolated horizons consistent, and completely general, \textit{ex post facto}.

Reinterpreting the quantization in this way, it is natural to consider different horizon states in the Hilbert space to represent different quantum shapes of the horizon.  (This view is also implicit in the previous work \cite{DMS:symh}.)  This reinterpretation of quantum states reconciles the entropy calculation of \cite{ABK:bhe} with the conceptual viewpoint that entropy arises by counting different microscopic shapes of the horizon, which is proposed, for example, in \cite{KR:fullbhe}.
Other recent work \cite{ENP:su2let, ENPP:su2} has reconciled the basic framework of \cite{ABK:bhe} with the key \textit{mathematical result} of \cite{KR:fullbhe} (see also \cite{LT:bhe}), that black hole states correspond to a single $SU(2)$ intertwiner.  This transpires by relaxing the $U(1)$ partial gauge-fixing at the horizon used in \cite{ABK:bhe}.  In contrast the present paper retains the assumption of that partial gauge fixing for simplicity.  A further synthesis of the results here and in \cite{ENP:su2let, ENPP:su2} would likely reconcile the frameworks of \cite{ABK:bhe} and \cite{KR:fullbhe} completely.

The following discussion is broken into short sections.  The first and second recall some useful facts regarding the canonical framework for classical isolated horizons.  The third derives the Chern--Simons boundary symplectic structure in the generic case.  The fourth examines quantization.  The fifth reinterprets the resulting quantum states, arguing that they represent distinct geometric shapes of the horizon.  The sixth and final section reassesses the statistical ensemble underlying the entropy calculation in this light. We conclude with
a summary.

\topic{Covariant Phase Space}

We use the first-order, Holst formulation \cite{H:act} of classical general relativity.  Its basic variables are a frame field $e^a_I$ and an independent connection $D_a$ in the frame bundle over spacetime $\sptm$.  The action of the theory is
\begin{equation}\label{actDef}
	S[e, D] := \frac{-1}{16\pi G}\, \int_{\sptm}\, \tr \Sigma \wedge \biggl( F + \frac{1}{\bip} \idual F \biggr),
\end{equation}
where $F_{abI}{}^J$ denotes the curvature of $D_a$, $\idual[0mu]$ denotes the Hodge dual in the internal space of components relative to the frame, the real constant $\bip$ is the \defn{Barbero--Immirzi parameter}, and
\begin{equation}\label{SigDef}
	\Sigma_{ab}{}^{IJ} := \idual[1mu] ( e^I \wedge e^J )_{ab} := e^K_a\, e^L_b\, \epsilon_{KL}{}^{IJ} .
\end{equation}
The classical equations of motion derived from (\ref{actDef}) imply that $D_a$ annihilates $e^a_I$, and that the spacetime metric $g_{ab} := \eta_{IJ}\, e^I_a\, e^J_b$ solves the Einstein equations.  Here, $\eta_{IJ}$ is the fixed, internal  Minkowski metric.

The \defn{covariant phase space} of the Holst formulation is the set of all solutions of its classical equations of motion.  The (pre-)symplectic structure on that phase space arises by integrating the \defn{symplectic current} 3-form
\begin{equation}\label{curMod}
	\biom(\delta_1, \delta_2)
		:= \frac{-1}{8\pi G} \tr \delta_{[1} \Sigma \wedge \biggl( \delta_{2]} D + \frac{1}{\bip} \idual[0mu] \delta_{2]} D \biggr)
\end{equation}
over a Cauchy surface $M$ in spacetime.  The brackets on the right here take the anti-symmetric part of the expression in two  perturbations $(\delta_{1, 2} \Sigma, \delta_{1, 2} D)$ of a given solution $(\Sigma, D)$ of the field equations.  The exterior derivative of $\biom(\delta_1, \delta_2)$ vanishes whenever both perturbations solve the linearized equations of motion on $(\Sigma, D)$, whence its integral over every Cauchy surface has the same value.

The motivation to use the Holst formulation classically comes from loop quantum gravity.  Fix a Cauchy surface $M$ and a unit time-like vector field $u^I$ in the internal space.  Imposing the \defn{time gauge} condition that $e^a_I\, u^I$ is the future-directed unit normal to $M$, one finds
\begin{equation}\label{ipsCur}
	\pback[M]{\biom(\delta_1, \delta_2)} 	= \frac{1}{4\pi G}\, \delta_{[1} \Sigma^i \wedge \delta_{2]} \biggl( K_i + \frac{1}{\bip}\, \Gamma_i \biggr)
		+ \frac{1}{4\pi G \bip}\, \pback[M]{\delta_{[1} e^I \wedge \delta_{2]} \bigl( D e_I \bigr)},
\end{equation}
where the arrows denote the pullback to the Cauchy surface, $K_I := \pback{D}\, u_I$ describes its extrinsic curvature, $\Sigma^I := \pback{\Sigma}^{IJ}\, u_J$ are the 2-forms dual to the triad induced thereon, $\pback{\ed e}{}^K =: \Gamma^I \wedge \pback{e}{}^J\, \epsilon_{IJ}{}^{KL}\, u_L$ defines the \defn{spin connection} $\Gamma^I$ of that triad, and lowercase indices denote orthogonality to $u^I$ in the internal space.  The second term in (\ref{ipsCur}) vanishes by the linearized equations of motion, while the first involves the \defn{Ashtekar--Barbero connection}
\begin{equation}\label{ipsBar}
	\biA_a^i := \Gamma_a^i + \beta K_a^i.
\end{equation}
This and the \defn{flux form} $\Sigma_{ab}^i$ are the basic variables used in the loop quantization of the theory \cite{AL:bisr}.

\topic{Horizon Boundary Conditions}

The spacetimes of interest here have an inner boundary $\Delta$ at an \defn{isolated horizon} \cite{Aea:ihLett, AFK:ham, ABL:mech, ABL:geom}.  Such horizons have the key features of a Killing horizon in a stationary spacetime needed to establish the laws of black hole mechanics, but do not demand global symmetry.  Let us collect a few facts from the precise definition that will be useful below.

An isolated horizon $\Delta$ is a null surface with topology $S^2 \times \Re$ and a preferred null normal $\ell^a$ fixed up to scaling by a \textit{constant}.  In addition, $\Delta$ admits a preferred foliation by 2-spheres $S_v$, with $v$ real, called its ``good cuts'' in analogy with null infinity \cite{ABL:geom}.  The only attribute of this geometrically preferred foliation we need below is merely that it exists.

We partially gauge fix the frame field $e^a_I$ at the horizon by fixing an internal null tetrad field $(\ell^I, n^I, m^I, \bar m^I)$ there and demanding
\begin{equation}\label{lnFix}
	e^a_I\, \ell^I \propto \ell^a \qquad\text{and}\qquad \pback{n_a} := \pback{e^I_a}\, n_I \propto \nu_a,		
\end{equation}
where the arrow denotes the pullback to $\Delta$, $\nu_a$ is the normal to $S_v$ within $\Delta$ such that $\ell^a\, \nu_a = -1$, and the proportionalities are \textit{constant} over the horizon.  In the covariant phase space framework below, we will restrict attention to Cauchy surfaces $M$ that intersect $\Delta$ in leaves $S_v$ of its preferred foliation.  When the time gauge condition described above is also imposed, we will further restrict the choice of the internal null tetrad at the horizon such that
\begin{equation}\label{rdef}
	\frac{\ell^I + n^I}{\sqrt{2}} = u^I, \qquad\text{whence}\qquad r^I := \frac{\ell^I - n^I}{\sqrt{2}}
\end{equation}
is a fixed, unit space-like vector orthogonal to $u^I$ in the internal space.  According to the definitions above, $r^I\, e^a_I$ is the unit normal to the inner boundary $S_v$ of $M$.  Thus, the only \textit{local} internal gauge freedom remaining in the frame field $e^a_I$ are the \defn{spin transformations} described by $m^a := m^I\, e^a_I \mapsto \ee^{2\ii\theta} m^a$ and its complex conjugate.  This reduces the internal gauge group to $U(1)$ at the horizon.

The null congruence generated by $\ell^a$ is required to be expansion- and shear-free.  This implies that there exist 1-form fields $\varpi_a$, $\varkappa_a$ and $\vartheta_a$ intrinsic to $\Delta$ such that
\begin{equation}\label{vkptDef}
		\pback{\grad\!{}_a}\, \ell^b =: \varpi_a\, \ell^b
			\qquad\text{and}\qquad \pback{\grad\!{}_a}\, \bar m^b =: \varkappa_a\, \ell^b + \ii\, \vartheta_a\, \bar m^b.
\end{equation}
Of the three, only $\varkappa_a$ is complex.  Under the local spin transformations described above, $\varpi_a$ is invariant, $\varkappa_a$ changes phase locally, and $\vartheta_a$ transforms as a $U(1)$ connection.  In fact, $\vartheta_a$ is just the intrinsic spin connection form on $S_v$.  Taking another derivative of (\ref{vkptDef}) leads to
\begin{alignat}{2}\label{vkCur}
	\ed\varpi		&= \bigl( \ii\, R_{abcd}\, \bar m^a\, m^b\, n^c\, \ell^d \bigr)\, \epsilon	 
					&&=: 2 \imag \Psi_2\, \epsilon, \\[1ex]\label{vtCur}
	\ed\vartheta	&= \bigl( R_{abcd}\, \bar m^a\, m^b\, \bar m^c\, m^d \bigr)\, \epsilon	
					&&=: \bigl( 2 \real \Psi_2 - m^a\, \bar m^b\, R_{ab} - \tfrac{1}{6} R \bigr)\, \epsilon = -\tfrac{1}{2} \sR\, \epsilon,
\end{alignat}
where $\epsilon := \ii\, m \wedge \bar m$ and $\sR$ denote the area element and \textit{intrinsic} scalar curvature on $S_v$, respectively, and the Newman--Penrose scalar $\Psi_2$ may be evaluated in \textit{any} null tetrad containing $\ell^a$.  Finally, (\ref{vkptDef}) shows that the flow generated by $\ell^a$ preserves the (degenerate) metric $q_{ab} := \smash{\pback{g_{ab}}}$ induced on $\Delta$.  This metric symmetry is required to extend to the horizon connection, whence $\Lie_\ell\, \varpi$, $\Lie_\ell\, \varkappa$, and $\Lie_\ell\, \ed\vartheta$ all vanish.

\topic{Horizon Symplectic Structure}

A spacetime $\sptm$ with inner boundary $\Delta$ admits only \textit{partial} Cauchy surfaces $M$, which terminate at their intersections $S$ with that boundary.  Although the symplectic current (\ref{curMod}) remains closed, its integrals over a pair of partial Cauchy surfaces $M_\pm$ generally differ by its integral over the region of $\Delta$ between their inner boundaries $S_\pm$.  If, however, the pullback $\pback{\biom}{(\delta_1, \delta_2)}$ to $\Delta$ is an exact exterior derivative, say $\ed\varsigma(\delta_1, \delta_2)$, then the integral over $\Delta$ reduces to a difference of integrals over $S_\pm$.  It follows that
\begin{equation}\label{bsInt}
	\biOm_{(M_\pm)}(\delta_1, \delta_2) := \int_{M_\pm} \biom(\delta_1,\delta_2) - \oint_{S_\pm} \varsigma(\delta_1, \delta_2)
\end{equation}
has the same value for any two partial Cauchy slices $M_\pm$.  The sign in the second term, the \defn{surface symplectic structure}, arises because the orientation of $S_\pm$ as the inner boundary of $M_\pm$ assumed here is opposite to its orientation as the boundary of a portion of $\Delta$.

Our phase space consists of spacetimes $\sptm$ that (a) contain a single isolated horizon $\Delta$ as their inner boundary, with (b) a fixed, but arbitrary, total area $\ad$.  Finally, as mentioned in the previous section, we also restrict (c) to partial Cauchy surfaces $M_\pm$ whose inner boundaries $S_\pm$ are leaves $S_v$ of the preferred foliation (\textit{i.e.}, ``good cuts'') of $\Delta$.  Then, the pullback $\pback{\biom}(\delta_1, \delta_2)$ to $\Delta$ has the form \cite{AK:idh}
\begin{equation}\label{pbMod2}
	\pback{\biom}(\delta_1, \delta_2) = \frac{1}{4\pi G}\, \delta_{[1} \epsilon \wedge \delta_{2]} \varpi
		+ \frac{1}{4\pi G \beta}\, \delta_{[1} \epsilon \wedge \delta_{2]} \vartheta.
\end{equation}
Moreover, the first term here is exact and the resulting integrals over $S_\pm$ vanish due to the restrictions (b) and (c) above.

The second term in (\ref{pbMod2}) simplifies if we introduce a \defn{curvature potential} $\psi$ on $S_v$ such that
\begin{equation}\label{cpDef}
	\Lap\psi := \sdual\ed\sdual\ed\psi := \sR - \expect{\sR} \qquad\text{and}\qquad \expect{\psi} := 0.
\end{equation}
Here, $\Lap$ denotes the Hodge Laplacian, $\sdual$ the Hodge dual, $\sR$ the scalar curvature, all intrinsic to $S_v$, and $\expect{f}$ denotes an average over $S_v$.  That is, $\expect{f}$ is the integral of $f \epsilon$ over $S_v$ divided by $\ad$.  The averaged term on the right side of the first equation is necessary for $\psi$ to exist, while the second condition makes $\psi$ unique.  We then define
\begin{equation}\label{rvtDef}
	\ring V_a := \tfrac{1}{2} \vartheta_a + \tfrac{1}{4} (\sdual \ed\psi)_a
		:= \tfrac{1}{2}\, \vartheta_a - \tfrac{1}{4}\, \epsilon_{ab}\, \hat q^{bc}\, \grad\!{}_c \psi,
\end{equation}
where $\hat q^{bc}$ denotes the inverse metric on $S_v$.  The factor of one half here helps to compare to a connection defined in \cite{AEV:qaxi}, also denoted $\ring V$.  Note, however, that (\ref{rvtDef}) defines $\ring V$ on \textit{any} horizon, regardless of symmetry. Also note that the second term in (\ref{rvtDef}) is invariant under local spin transformations of $m^a$ since $\psi$ arises directly from the metric.  It follows that $\ring V$ transforms like a $U(1)$ connection, albeit with half the weight of $\vartheta$.  Furthermore, (\ref{vtCur}) gives
\begin{equation}\label{rvtCur}
	\ed\ring V = \frac{1}{2}\, \ed\vartheta + \frac{1}{4}\, \ed\sdual\ed\psi
		= -\frac{\sR}{4}\, \epsilon + \frac{1}{4}\, \Lap\psi\, \epsilon
		= -\frac{\expect{\sR}}{4}\, \epsilon
		= -\frac{2\pi}{\ad}\, \ppback{\Sigma}_i\, r^i,
\end{equation}
where the double arrow in the last expression denotes the pullback to $S$ and $r^i$ is introduced in (\ref{rdef}).  Note that the area \textit{element} on $S$ is naturally a \textit{pseudo}-form, whereas the final expression in (\ref{rvtCur}), which we call the \defn{area flux} below, is naturally a form.  The left side is certainly a form, so there is already a reason at the classical level to prefer the area flux to the area element in this expression.  This becomes a crucial distinction in the quantum theory.

Contracting $\ell^a$ with (\ref{rvtDef}) causes the second term on the right to vanish.  This determines the only component of $\vartheta_a$ that enters (\ref{pbMod2}) in terms of $\ring V_a$.  Together, these observations yield
\begin{equation}\label{pbBar}
	\delta_{[1} \epsilon \wedge \delta_{2]} \vartheta
        = \delta_{[1} \biggl(-\frac{4}{\expect{\sR}} \, \ed \ring V \biggr)\wedge
        \delta_{2]} \bigl(2\ring V\bigr)
        = - \frac{4}{\expect{\sR}}\, \ed (\delta_1 \ring V \wedge \delta_2 \ring V).
\end{equation}
With $\ad$ fixed, $\expect{\sR} = 8\pi / \ad$ is a \textit{topological} invariant by the Gauss--Bonnet theorem.  It therefore passes through the variation to give the above equation.

Substituting (\ref{pbBar}) into (\ref{pbMod2}), and the result into (\ref{bsInt}), we find the symplectic structure
\begin{equation}\label{curInt}
	\biOm(\delta_1, \delta_2) = \frac{1}{4\pi G \bip} \int_M \delta_{[1}
        \Sigma^i \wedge \delta_{2]} \biA_i
		+ \frac{1}{2\pi} \frac{\ad}{4\pi G \bip} \oint_S \delta_1 \ring V \wedge \delta_2 \ring V.
\end{equation}
Here, $\biOm_{\mathrm{B}}(\delta_1, \delta_2)$ denotes the integral of (\ref{curMod}) over a partial Cauchy surface $M$ with inner boundary $S = S_v$.  This result has precisely the general form found in symmetric cases \cite{ABCK:bhe, ACK:ihps, ABK:bhe, AEV:qaxi}, but holds throughout the \textit{full} classical phase space of \textit{all} isolated horizons with area $\ad$.

This full classical phase space contains subspaces corresponding to the symmetric cases studied previously.  We check that (\ref{curInt}) reproduces the known symplectic structures on those subspaces by solving (\ref{cpDef}) explicitly.  This is immediate in spherical symmetry, where $\psi = 0$.  It follows that $\ring V = \vartheta / 2$ and, indeed, the spherically symmetric symplectic structure includes a Chern--Simons horizon term for precisely this part of the physical spin connection \cite{ABCK:bhe, ABK:bhe}.  In axisymmetry, we work in the canonical coordinates $(\zeta, \phi)$ of \cite{AEPV:mmih}, which generalize the coordinates $(\cos\theta, \phi)$ on a round sphere.  The potential $\psi$ itself is axisymmetric in this case, and (\ref{cpDef}) becomes
\begin{equation}\label{asLap}
	- \frac{4\pi}{\ad}\, \frac{\partial\ }{\partial\zeta} \bigl( f(\zeta)\, \psi'(\zeta) \bigr) = - \frac{4\pi}{\ad}\, f''(\zeta) - \frac{8\pi}{\ad},
\end{equation}
where primes denote derivatives and $f(\zeta)$ is the squared norm of the axial Killing field $\varphi^a = \partial_\phi^a$.  The constant of integration in the first integral of this expression must vanish to avoid logarithmic singularities in $\psi(\zeta)$ at the poles where $\varphi^a$ vanishes.  Thus, we find
\begin{equation}\label{asRing}
	\ring V = \tfrac{1}{2}\, \vartheta + \tfrac{1}{4}\, \sdual\ed\psi = \tfrac{1}{2}\, \vartheta - \tfrac{1}{4}\, \bigl( f'(\zeta) + 2\zeta \bigr)\, \ed\phi.
\end{equation}
This is precisely the axisymmetric formula \cite{AEV:qaxi}.  All previous expressions of the horizon symplectic structure therefore represent \textit{pullbacks} of (\ref{curInt}) to subspaces of the full phase space of all horizons.

\topic{Quantization}

The symplectic structure (\ref{curInt}) is almost identical to that studied in \cite{ABCK:bhe, ACK:ihps, ABK:bhe}.  The only actual difference is that its horizon term involves $\ring V$ instead of $\vartheta / 2$.  We now argue that in fact \textit{all} elements of the classical framework necessary for quantization in \cite{ABCK:bhe, ACK:ihps, ABK:bhe} are also present here.  Moreover, the only element \textit{not} present, the restriction to round horizon geometries, was never truly incorporated at the quantum level in \cite{ABCK:bhe, ACK:ihps, ABK:bhe}.  In this sense, we claim that the quantization described in detail in \cite{ABK:bhe} is properly viewed as that of the \textit{full} phase space of generic isolated horizons defined in this paper.  Let us proceed by first recalling the key steps of the detailed quantization, confirming that each remains viable here.

The sum of bulk and surface terms in the the classical symplectic structure (\ref{curInt}) suggests that initially the Hilbert space of the quantum theory should be a tensor product $\Hil[^pre] = \Hil[_B] \otimes \Hil[_S]$ of bulk and surface factors.  These consist of suitable functions of a generalized $SU(2)$ connection on $M$, usually denoted $A$, and of a generalized $U(1)$ connection on $S$, which we shall denote $X$, respectively.  We use the non-prejudicial notation $X$ for the quantum $U(1)$ Chern--Simons connection to emphasize a point.  Namely, we mean to expropriate the key results of \cite{ABK:bhe}, but the \textit{classical} horizon connection $\ring V$ we use here is certainly \textit{not} the same as the intrinsic spin connection $\vartheta / 2$ on the horizon used there.  We call the quantum horizon connection $X$ at this stage to emphasize that it currently has no interpretation in terms of \textit{spatial geometry}.  Indeed, $\Hil[^pre]$ is simply a tensor product in which the surface and bulk components of a state are fully independent, and \textit{all} operators encoding the spatial geometry are already represented on $\Hil[_B]$ alone.

We now recall how the boundary and bulk degrees of freedom become related in the quantum theory.  Already in the classical theory, $\ring V$ may be characterized as the unique $U(1)$ connection on $S$ whose holonomy along a given oriented loop $C$ is
\begin{equation}\label{rvtHol}
	U[\ring V, C] := \exp \oint_C \ii \ring V
		= \exp \biggl( -\frac{2\pi\ii}{\ad} \int_{\Int C} \Sigma_i\, r^i \biggr),
\end{equation}
the \defn{interior} $\Int C$ being selected by the orientation $S$.  This formula, or its infinitesimal version (\ref{rvtCur}), shows that the \textit{gauge-invariant part} of $\ring V$ is \textit{completely} and \textit{locally} determined by the frame field at $S$.  That is, even though inverting the Laplacian in (\ref{cpDef}) renders $\ring V$ in (\ref{rvtDef}) \textit{non-local} in the geometric spin connection $\vartheta / 2$ intrinsic to $S$, its gauge-invariant content is nonetheless \textit{locally} determined by the full set of basic variables of the theory.  Note that $\ring V$ is local in $\vartheta / 2$ if and only if the horizon is spherically symmetric, and then the two are equal.

The kinematical Hilbert space $\Hil[^kin]$ arises in the present case by elevating (\ref{rvtHol}) to an operator equation on $\Hil[^pre]$, the two sides acting in different factors of the tensor product.  This defines the \defn{quantum horizon boundary condition}
\begin{equation}\label{qhbc}
	\hat I_{\mathrm{B}} \otimes \hat U[X, C]
		= \exp \biggl( - \frac{2\pi\ii}{\ad}\, \hat\Sigma[\Int C, r] \biggr) \otimes \hat I_{\mathrm{S}},
\end{equation}
which plays two pivotal roles in the quantization.  First, it fixes the Chern--Simons boundary state in $\Hil[_S]$ in terms of data on edges that terminate at $S$ in a given spin network state in $\Hil[_B]$.  Thus, just as (\ref{rvtHol}) dictates $\ring V$ classically in terms of the bulk variables, so the quantum bulk state dictates the quantum boundary state.  Second, (\ref{qhbc}) links (quantum) $U(1)$ gauge transformations in $\Hil[_S]$ to $SU(2)$ gauge transformations in $\Hil[_B]$ that preserve the internal direction $r^i$ at $S$.  This guarantees gauge invariance of the overall state.  Again, there is a classical antecedent in that such a bulk gauge transformation dictates a $U(1)$ gauge transformation of $\vartheta$, and thus of $\ring V$ in (\ref{rvtDef}).

In addition to fixing the boundary state in terms of the bulk state, imposing (\ref{qhbc}) in $\Hil[^pre]$ does restrict the bulk state itself, but only slightly.  In particular, the bulk state must lie in the subspace of $\Hil[_B]$ spanned by eigenstates of $\hat\Sigma[S, r]$ whose eigenvalues are integer multiples of $\ad$.  These are spin network states that satisfy
\begin{equation}\label{csrev}
	\sum_p 8 \pi G \bip m_p = n \ad \qquad\Leftrightarrow\qquad
		\sum_p 2 m_p = 0 \mod k := \frac{\ad}{4 \pi G \bip},
\end{equation}
where $p$ runs over the horizon punctures.  This ensures that holonomies of the $U(1)$ connection $X$ are defined consistently in the quantum Chern--Simons theory in the sense that reversing the orientation $C \mapsto -C$ of the loop replaces the holonomy on the left side of (\ref{qhbc}) with its Hermitian adjoint.


Solving the diffeomorphism constraint for generic horizons proceeds exactly as described in \cite{ABK:bhe}.  Both the definition (\ref{cpDef}) of $\psi$ and thus the definition (\ref{rvtDef}) of $\ring V$ are manifestly diffeomorphism covariant.  Thus, a bulk diffeomorphism preserving $S$ induces the same transformation of $\ring V$ on a generic horizon that it does of $\vartheta / 2$ in spherical symmetry.  The passage to the Hilbert space $\Hil[^diff]$ of solutions of the diffeomorphism constraint is therefore unchanged.  The subsequent passage to the physical Hilbert space $\Hil[^phys]$ of solutions of the Hamiltonian constraint is even simpler because, as in \cite{ABK:bhe}, the lapse function must vanish at $S$.  The resulting constraints do not affect surface states.

We now make the case that $\Hil[^phys]$, as constructed in \cite{ABK:bhe}, does not incorporate spherical symmetry.  We first mention a simple, but heuristic, argument based on the central result of \cite{ABK:bhe}, that $\Hil[^phys]$ supports an ensemble of states with (horizon) entropy $\Sd = \ad / 4 G \hbar$.  This diverges as $\hbar \to 0$, implying that $\Hil[^phys]$ comprises infinitely many semi-classical horizon states.  In contrast, the phase space of spherically symmetric horizons includes, modulo diffeomorphisms, exactly one such state.  Thus, it appears that $\Hil[^phys]$ contains semi-classical states describing non-spherical horizons.

Now let's consider the question in the full quantum theory, not just in its semi-classical limit.  Classically, spherical symmetry restricts not only the $U(1)$ horizon connection, but also the \textit{bulk} frame field at every point of the horizon surface, anticipating a similar \textit{local} restriction of the bulk quantum state.  But the only such restriction, beyond the usual diffeomorphism and Hamiltonian constraints, is (\ref{csrev}).  This single, \textit{global} restriction is nothing like a \textit{local}, classical symmetry condition.  A definitive resolution of the matter would begin with a detailed study of the operators on $\Hil[^kin]$ corresponding to the classical bulk degrees of freedom constrained by spherical symmetry.  These are the flux operators $\hat\Sigma^i$, the quantum analogues of the classical triad $e^i$.  The next section is devoted to this issue.

We conclude this section with a comment.  As emphasized above, the quantum $U(1)$ Chern--Simons connection $X$ is unrelated to spatial geometry in $\Hil[^pre]$.  It is \textit{only} the quantum horizon boundary condition (\ref{qhbc}) that gives $X$ its physical interpretation.  Because that operator equation mirrors (\ref{rvtHol}), the quantum theory \textit{forces} us to interpret $X$ as nothing but the quantum analogue of the ``area flux connection'' $\ring V$.  That is, the quantum horizon boundary condition (\ref{qhbc}) is identical to that in \cite{ABK:bhe}, and one may consider (\ref{rvtHol}) its proper classical antecedent.  In particular, this implies that $X$ cannot be considered a quantum analogue of the classical spin connection on the horizon, and its holonomies do not generally measure quantum deficit angles.

\topic{Quantum Horizon Geometry}

The preceding section has argued that the kinematical Hilbert space $\Hil[^kin]$ constructed in \cite{ABK:bhe} actually describes \textit{all} isolated horizons, regardless of shape. This section explores how, in principle, one would go about measuring the quantum geometry of the horizon surface in a particular quantum state in $\Hil[^kin]$.  These comments further clarify our claim that no symmetry is present quantum mechanically.

\newcommand\at{\text{ at }}
Quite generally in loop quantum gravity, the geometrical content of states is measured using the smeared flux operators
\begin{equation}\label{fluxOp}
	\frac{1}{8\pi G \bip} \hat\Sigma[T, f] := \frac{\hbar}{2} \sum_{x \in T} f^i(x) \sum_{e \at x} \kappa(T, e)\, \hat J^{(x, e)}_i,
\end{equation}
where is a $T$ transversely oriented 2-surface in $M$ and the sums are over all points $x$ of $T$ and all analytic curves $e$ extending from each $x$.  The classical analogues of these operators are integrals of the smeared flux $f^i\, \Sigma_i$ over $T$.  We define all three factors in the summand of (\ref{fluxOp}) exactly as in \cite{AL:bisr}.  In particular, $\hat J_i^{(x, e)}$ generates either left- or right-translations, depending on whether $e$ starts or ends at $x$, of the generalized holonomy of the Ashtekar--Barbero connection along $e$.  In addition, $\kappa(e, T)$ vanishes if $e$ does not intersect or lies entirely within $T$, and is $\pm 1$ if $e$ extends either above or below $T$ in the transverse orientation of the surface, respectively.  Finally, $f^i(x)$ is simply a smearing function and the factor on the left arises from the coefficient of the bulk term in the symplectic structure (\ref{curInt}).  When acting on a spin network state, (\ref{fluxOp}) reduces to a finite sum over vertices and edges in the network.

The fluxes (\ref{fluxOp}) are naturally conjugate to the $SU(2)$ holonomy operators $\hat U[\biA, C]$ of the generalized Ashtekar--Barbero connection.  After quantization, however, \cite{ABCK:bhe, ACK:ihps, ABK:bhe} consider holonomies only along curves $C$ \textit{transverse} to $S$, and thus only spin network states without edges tangent to the horizon.  The rationale for this is evident classically, where the pullback to $S$ of the extrinsic curvature $K_{ab}$ of $M$ is
\begin{equation}\label{extCur}
	\ppback{K_{ab}} = k_{ab} = \real [\varkappa_a\, m_b]\, \ell^c\, r_c.
\end{equation}
Here, $r_c$ and $k_{ab}$ denote the unit normal and extrinsic curvature of $S$ within $M$, respectively, and $\varkappa_a$ is defined in (\ref{vkptDef}).  Note that $k_{ab}$ is determined by the spin connection $\Gamma_a^i$ of the triad on $M$.  Thus, only the transverse part $r^a\, \biA_a^i$ of the Ashtekar--Barbero connection (\ref{ipsBar}) retains degrees of freedom independent of the triad field at the horizon.  It is therefore natural to restrict to curves whose holonomies probe those degrees of freedom.  Furthermore, the \defn{pullback fluxes} $\hat\Sigma[T, f]$, where $T$ is a region within the horizon surface $S$, already provide natural conjugates to such \defn{transverse holonomies}.  The algebra on $\Hil[_B]$ of transverse holonomies and pullback fluxes at the horizon plays a central role in the quantization scheme.

The point we wish to emphasize is that pullback fluxes and transverse holonomies form only \textit{part} of the full operator algebra available at the horizon.  Even excluding tangential holonomies, there are fluxes $\hat\Sigma[T, f]$ through surfaces $T$ intersecting $S$ \textit{transversely} to consider.  In the classical theory, such \defn{transverse fluxes} reveal the \textit{intrinsic} horizon geometry because the transverse component $r^a\, \Sigma_{ab i} = \epsilon_{bc}\, e^c_i$ of the classical flux encodes the dyad on $S$ from which the horizon metric $q_{ab}$ derives.  In contrast, the pullback flux $\ppback{\Sigma_{ab}}{}_i = \epsilon_{ab}\, r_i$ reveals only the classical area element, which roughly speaking gives only the determinant of $q_{ab}$.  Thus, transverse fluxes are essential to analyze the classical horizon geometry.

It is also clear in the full quantum theory that pullback fluxes alone do not suffice to extract all information available at the horizon in a particular quantum state.  The generators $\hat J_i^{(x, e)}$ in the sum (\ref{fluxOp}) associated with different curves $e$ emanating from a given point $x$ of the horizon are \textit{independent}.  All of these generators contribute to a pullback flux $\hat\Sigma[T \subset S, f]$ with the same sign $\kappa(T \subset S, e)$ because $S$ is the boundary of $M$.  Thus, the pullback flux reveals only the \textit{sum} of the $\hat J_i^{(x, e)}$ over all $e$ at each $x$, and not the generators individually.  In contrast, a given $e$ may fall on either side of a transverse surface $T$, and thus may enter (\ref{fluxOp}) with either sign.  This flexibility allows one to recover generators from fluxes.  The only way to recover all of the information probed by the generators is therefore to employ all of the fluxes at the horizon.

With this understanding of the operators available at the horizon, one can now imagine how in principle to explore its intrinsic quantum geometry.  For example, the generators, or equivalently the transverse fluxes, should enable one to define operators corresponding to the lengths of curves within the horizon surface \cite{T:len, B:len}.  We do not attempt to write down such operators explicitly in this paper, nor to analyze the full operator algebra at the horizon in detail.  Rather, our purpose is merely to articulate a point of view that the \textit{intrinsic quantum geometry} of the horizon inheres in a part of the operator algebra that heretofore has not been studied thoroughly in the literature, and which certainly is not explicitly constrained by symmetry at the quantum level in \cite{ABK:bhe}.

Let us conclude with two comments.  First, the transverse flux operators highlighted above \textit{already exist} on $\Hil[^kin]$ in the standard quantization.  The boundary conditions do not \textit{create}, but rather \textit{eliminate}, independent degrees of freedom at the horizon.  The remaining data are parametrized using the fluxes and transverse holonomies.  Operators representing the tangential holonomies could in principle be constructed from these.  In this sense, all of the usual gravitational degrees of freedom at the horizon are represented on $\Hil[^kin]$.

Second, the view of quantum horizon geometry adopted here is quite different from that in previous works.  In particular, the operators used to describe quantum horizon geometry in \cite{ABCK:bhe, ACK:ihps, ABK:bhe, AEV:qaxi} are the area and the area flux, which are functions only of pullback fluxes.  This is possible classically only if the shape of the horizon is fixed up to diffeomorphism using symmetry.  We have argued that the symmetry does not persist after quantization and that, if the horizon shape is not fixed classically, then it is expressed in the \textit{transverse} fluxes.  Finally, we have shown explicitly that the quantum analogues $\hat\Sigma[T \not\subset S, f]$ of the transverse fluxes are independent of the pullback fluxes $\hat\Sigma[T \subset S, f]$ as operators on $\Hil[^kin]$.  Taken together, these facts indicate the need for a broader interpretation of intrinsic quantum horizon geometry not based purely on pullback fluxes.  Our proposal is essentially to use the standard formulation of loop quantum geometry in terms of \textit{all} fluxes available at the horizon.

\topic{Ensemble and Entropy}

Quite generally, the entropy of a physical system depends on one or more macroscopic parameters.  Fixing values of these parameters determines a microcanonical ensemble of states in which the corresponding macroscopic observables take those values.  This ensemble typically contains many states that differ in microscopic details, but give the same desired macroscopic results.  The entropy for given parameter values is a measure of how many states are in the ensemble.

We emphasize these points to contrast an interpretation of the entropy calculation \cite{ABK:bhe} advocated below with that previously ascribed to it in \cite{ABCK:bhe, ACK:ihps, ABK:bhe, AEV:qaxi}.  To be clear, we propose no modification to the detailed mathematical calculation, only to its physical interpretation.  At issue is the number and nature of parameters used to select the ensemble.

Physically, the ensemble certainly ought to fix the total horizon area and perhaps a few other macroscopic charges of the black hole.  But this is not what is done in \cite{ABCK:bhe, ACK:ihps, ABK:bhe, AEV:qaxi}.  The situation is clearest in the axisymmetric case \cite{AEV:qaxi}, where the ensemble ostensibly fixes the horizon area as well as an \textit{infinite} collection of certain multipole moments \cite{AEPV:mmih} of its geometry.  Classically, the complete collection of multipole moments characterize an axisymmetric horizon geometry up to diffeomorphism, fixing details of that geometry down to the very shortest length scales.  There is nothing \textit{macro}\-scopic about fixing \textit{all} of the multipole moments.  The situation is less transparent in spherical symmetry because \cite{ABCK:bhe, ACK:ihps, ABK:bhe} do not \textit{explicitly} restrict the ensemble.  Rather, it is held (erroneously in our view) that $\Hil[^phys]$ describes nothing but round horizons, and thus the ensemble as well.  But again, asserting the symmetry classically at all length scales clearly fixes degrees of freedom that physically can only be considered \textit{micro}\-scopic.

We have argued above that the Hilbert space $\Hil[^phys]$ defined in \cite{ABK:bhe} is properly interpreted to comprise quantum states corresponding to \textit{all} horizons of total area $\ad$, regardless of shape.  Under this interpretation, there is no obstacle to defining an ensemble of states within $\Hil[^phys]$ fixing only the manifestly \textit{macro}scopic parameter $\ad$.  This ensemble happens to coincide precisely with that defined in \cite{ABK:bhe}.  It contains all eigenstates of the total area operator $\ad[\hat]$ whose eigenvalues are within a certain tolerance $\delta$ of the classical value $\ad$ prior to quantization.  The horizon entropy arises by counting just the distinct surface states in this ensemble because the bulk states describe field excitations elsewhere in spacetime.

Intuitively, the entropy in this interpretation measures the variety of \textit{shapes} a horizon with given area $\ad$ may have.  Moreover, quantization \textit{cuts off} geometric degrees of freedom in this interpretation, rather than appearing to multiply them.  More precisely, the interpretation of \cite{ABCK:bhe, ACK:ihps, ABK:bhe, AEV:qaxi} holds that quantizing a phase space comprising (up to diffeomorphism) a single horizon state, wherein any reasonable horizon entropy ought to vanish, yields a Hilbert space in which the entropy is non-zero.  In contrast, our interpretation holds that quantizing a phase space comprising a continuous infinity of horizon geometries, wherein the entropy ought to diverge, yields a Hilbert space in which the entropy is finite.  The latter interpretation is of course quite common in quantum mechanics, with historical roots going back to the very foundation of the theory.

This discussion has sidestepped the important issue of how the horizon states are actually to be counted.  The literature contains (at least) two schemes for doing this, and in principle our proposal is compatible with both.  One scheme, the one adopted in \cite{ABCK:bhe, ACK:ihps, ABK:bhe, AEV:qaxi}, counts exactly the states of the $U(1)$ Chern--Simons theory.  This is sometimes called \defn{$m$-counting} because it counts distinct labelings by magnetic quantum numbers (\textit{i.e.}, pullback fluxes) of the horizon punctures.  The alternative, called \defn{$jm$-counting} \cite{GM:jm}, is to count distinct labelings of the punctures by both spin \textit{and} magnetic quantum numbers.  The two schemes differ only in the value of the Barbero--Immirzi parameter they select \cite{ABDFV,DL,Meissner}, and it is not clear which one is ``correct.''  The inclusion of generic isolated horizons achieved here adds nothing to this discussion.  However, we do remark that this question seems connected to recent efforts \cite{ENP:su2let, ENPP:su2} to quantize isolated horizons without reducing the $SU(2)$ gauge to $U(1)$ at the horizon.

\section{Conclusions}

At the classical level, the key technical result of this paper is that the symplectic structure (\ref{curInt}) is conserved on the \textit{full} phase space of \textit{all} isolated horizons of area $\ad$.  The proof depends critically on the definition (\ref{rvtDef}) of a new $U(1)$ connection $\ring V$ on the horizon.  It is worthwhile to note two things about this definition.  First, $\ring V$ yields a Chern--Simons surface symplectic structure (\ref{curInt}) not only on the phase space of horizons of a fixed, but arbitrary, shape, but permits one to fit all of those phase spaces together into a single, overall phase space in which the shape may vary.  This is a stark contrast to past work \cite{ABCK:bhe, ACK:ihps, ABK:bhe, AEV:qaxi}, wherein the horizon shape had to be fixed classically in order to construct a conserved symplectic structure.  Second, the formula (\ref{rvtHol}) for the classical holonomies of $\ring V$ parallels \textit{exactly} the quantum horizon boundary condition (\ref{qhbc}) at the heart of the seminal quantization \cite{ABK:bhe} of isolated horizons.  One could say that the importance of an ``area flux connection'' whose curvature is a flux that is \textit{conjugate} to the usual Ashtekar--Lewandowski connection was first perceived in \cite{ABCK:bhe, ACK:ihps, ABK:bhe} at the \textit{quantum} level, and that what we have done here is to introduce a \textit{classical} connection with the same geometric meaning on a general horizon.

At the quantum level, the key technical result of this paper is that each step of the detailed quantization \cite{ABK:bhe} of isolated horizons remains viable if it begins with the phase space of classical horizons of variable shape.  This is surprising because it seems as though two phase spaces, one restricted to spherical symmetry and considerably smaller than the other, lead to identical Hilbert spaces.  However, one can see directly in the final quantum theory that the geometric degrees of freedom probed by fluxes at the horizon are not restricted beyond the usual constraints of general relativity.  This would not be the case if symmetry were present.  Thus the symmetry present classically in \cite{ABCK:bhe, ACK:ihps, ABK:bhe, AEV:qaxi} was not reasserted after quantization and the final Hilbert space properly corresponds to the \textit{full} phase space in which all horizon shapes are included.  Indeed, arguably the key achievement of this paper has been precisely to introduce
a classical framework that exactly matches the quantum framework in \cite{ABCK:bhe, ACK:ihps, ABK:bhe, AEV:qaxi}.

These technical results shed new light on the calculation of black hole entropy in loop quantum gravity.  Neither the entropy calculation nor the predicted value(s) of the Barbero--Immirzi parameter change, but how one interprets the entropy physically does.  Namely, the absence of symmetry at the quantum level implies that the states being counted to find the entropy correspond to different quantum shapes of the horizon.  This interpretation of the statistical entropy of black holes in loop quantum gravity strongly resembles familiar interpretations of more ordinary statistical mechanical systems.

\ack

We thank Carlo Rovelli for discussions, Thomas Thiemann for remarks on a prior draft of this paper, and Abhay Ashtekar for encouraging us to finish this work.  J.E. was supported in part by NSF grant OISE 0601844 and the Alexander von Humboldt foundation.

\bibpage

\end{document}